\documentclass[12pt,letterpaper]{article}
\begin{document}

\newcommand{\be}{\begin{equation}}
\newcommand{\ee}{\end{equation}}

\author{H.~Neuberger\\ [7mm]
  {\normalsize\it Department of Physics and Astronomy, Rutgers University}\\
  {\normalsize\it Piscataway, NJ 08855, U.S.A} }

\title{
Complex Burgers' equation in 2D SU(N) YM.}
\maketitle \vskip 1.5cm

\abstract {\noindent An integro-differential equation satisfied by 
an eigenvalue density defined as the logarithmic derivative of the average inverse characteristic polynomial of a Wilson loop in two dimensional pure Yang Mills theory with 
gauge group SU(N) is derived from two 
associated complex Burgers' equations, with 
viscosity given by 1/(2N). The Wilson loop does not intersect itself and Euclidean
space-time is assumed flat and infinite. This result 
provides an extension of the infinite N 
solution of Durhuus and Olesen 
to finite N, but this extension is not unique.}

\bigskip
\newpage

\tableofcontents

\vskip 1cm

\section{Introduction.}

In~\cite{first}, in the context of two dimensional YM theory on the infinite plane with gauge group $SU(N)$, the function $\phi_N(y,\tau)$ defined by
\be
\phi_N(y,\tau)=-\frac{1}{N} \frac{\partial}{\partial y} \log \left [
e^{\frac{N}{2}\left ( \frac{\tau}{4} -y\right )} \langle \det (e^y +W)\rangle \right ]
\ee
was shown to satisfy Burgers' equation
\be
\frac{\partial{\phi_N}}{\partial \tau} +\phi_N \frac{\partial \phi_N}{\partial y} =\frac{1}{2N} \frac{\partial^2 \phi_N}{\partial y^2}
\label{burgersa}
\ee 
with initial condition
\be
\phi_N(y,0)=-\frac{1}{2}\tanh\frac{y}{2}
\ee
 $W$ is a Wilson operator associated with a non-selfintersecting
loop. $\tau$ measures the area enclosed by the loop in units of the
't Hooft gauge coupling. $\langle ...\rangle$ denotes 
averaging with respect to the exponent of the two dimensional YM action.
$\tau \ge 0$ and $y$ and $\phi_N$ are real. 

At $N=\infty$ a shock appears at $\tau=4$; for finite $N$ the shock is smoothed in a universal way in the regime $y\sim 0, \tau \sim 4$. $\phi_N$ admits a pole expansion with exactly integrable pole dynamics in a ``time'' $\tau$. 

The $N=\infty$ critical value $\tau=4$ corresponds to the Durhuus-Olesen (DO)~\cite{duol} phase transition point. That phase transition was found 
by solving a complex inviscid ($N=\infty$) Burgers' equation. More 
specifically, $y$ was taken to approach purely imaginary values 
and $\phi_N$ was complex. $\tau$ remained real and non-negative. 

The objective of this letter is to identify $\langle \det (z-W)^{-1} \rangle$ as an object that is more directly linked to the DO solution. In principle, there exist an arbitrary number of possible extensions
of the DO equation and solution; this choice is special because all finite $N$ effects are accounted for by a viscous term in an associated Burgers' equation with viscosity given by $\frac{1}{2N}$, similarly to~\cite{first}. The singularity structure for finite
$N$ is richer here than in~\cite{first}. 

\section{Conventions.}

The partition function of Euclidean 2D $SU(N)$ YM is written as:
\be
Z=\int [{\cal D} A_\mu ] e^{-\frac{1}{2g^2_{YM}} \int d^2 x tr F_{\mu\nu}
F_{\mu\nu}}
\ee
't Hooft's coupling is 
\be
\lambda=g^2_{YM} N
\ee
and the area enclosed by the loop is ${\cal A}$. 

The integration over $A_\mu$ at fixed $W$ induces a probability density for $W$, given by
\be
{\cal P}_N (W,\lambda {\cal A} ) =\sum_R d_R \chi_R (W)
e^{-\frac{\lambda {\cal A} }{2N} C_2 (R) } 
\ee
where $R$ denotes an irreducible representation of $SU(N)$ of
dimension $d_R$ and the character $\chi_R (W)$ satisfies:
\be 
\chi_R ({\bf 1}) = d_R,~~~\int dW \chi_R (W) \chi^*_S(W)=\delta_{RS}
\ee
$dW$ is the normalized Haar measure on $SU(N)$ and ${\cal P}_N$
is defined relative to it. $C_2(R)$ is the quadratic Casimir 
of $R$; for the defining representation $F$ we have:
\be
C_2 (F) = N-\frac{1}{N}
\ee
${\cal P}_N$ obeys:
\be
{\cal P}_N(W,\lambda{\cal A})={\cal P}_N(W^*,\lambda {\cal A})= {\cal P}^*_N(W,\lambda {\cal A})={\cal P}_N(W^\dagger,\lambda {\cal A})
\ee

The variable $\tau$ above is given by
\be
\tau=\lambda{\cal A}\left (1+\frac{1}{N}\right )
\ee
The probability density can be viewed as a function of $\tau$:
\be
P_N (W,\tau) ={\cal P}_N(W,\lambda{\cal A})
\ee

\section{Antisymmetric representations.~\cite{first}}

Expanding the characteristic polynomial we obtain a sum over
all $k$-fold antisymmetric representations $F^{\wedge k}$ of dimension 
$d_k={N \choose k}$:
\be
\det(z-W)=z^N\left [ 1+ \frac{1}{(-z)^N} +\sum_{k=1}^{N-1} \frac{\chi_k (W)}{(-z)^k}\right ]
\ee
For $W={\rm diag}(e^{i\theta_1},...,e^{i\theta_N})$ we have
\be
\chi_k (W)=\sum_{1\le j_1 < j_2....<j_k\le N} e^{i(\theta_{j_1} +\theta_{j_2}+....+\theta_{j_k})}
\ee
Also,
\be
C_2(k)=\frac{N+1}{N} k (N-k) = C_2(N-k)
\ee
For the computation of $\langle \det (z-W)\rangle$ only the representations $R=F^{\wedge k}$ in the sum giving $P_N$ contribute. 
One obtains, with real $y$:
\be
\langle \det(e^y +W)\rangle = e^{\frac{N}{2}\left (y-\frac{\tau}{4}\right )} \sum_{k=0}^N { N\choose k} 
e^{y\left ( k -\frac{N}{2}\right ) } 
e^{\frac{\tau}{2N}\left ( k -\frac{N}{2}\right )^2} 
\equiv e^{\frac{N}{2}\left (y-\frac{\tau}{4}\right )} q_N(y)
\ee

The forms of the prefactor of $y$ in the exponent and the prefactor of
$\tau$ in the exponent show that $q_N(y)$ obeys the (linear) heat equation
\be
\frac{\partial q_N}{\partial \tau} =
\frac{1}{2N} \frac{\partial^2 q_N}{\partial y^2} 
\label{heateq}
\ee
which leads to~(\ref{burgersa}).

\section{Symmetric representations.}

Equation~(\ref{heateq}) holds as a consequence of the linearity
of $P_N$ in representation space, the restriction of contributions
to a subset $R_k$ that can be labeled by an index $k$, and the quadratic
dependence of $C_2 (R_k) $ on $k$. The dimensions $d_k$ only enter
through the initial condition. Therefore, one 
expects a similar equation  
to hold for a sum over all $k$-fold symmetric representations. 
This time the range of $k$ extends to infinity, the generating function
of all the $R_k$ is no longer a polynomial and, consequently, a richer
analytic structure is expected. 

The generating function for all symmetric representations is the inverse
of the characteristic polynomial. The general formula
can be obtained by considering a diagonal $W$:
\be
\frac{1}{\det(z-W)} = \frac{1}{z^N} \prod_{j=1}^N \left [ \sum_{n=0}^N z^{-n} e^{in\theta_j}\right ] = \frac{1}{z^N}\left [ 1+\sum_{k=1}^\infty \frac{\chi_k (W)}{z^k}\right ],
\ee
where the character and dimension of each $R_k$ are given by:
\be
\chi_k (W)=\sum_{n_1,n_2,...,n_N \ge 0,~\sum_{j=1}^Nn_j =k }
e^{in_1\theta_1+in_2\theta_2+....+in_k\theta_k};~~d_k={{N+k-1} \choose {N-1}}
\ee
Most importantly, the second order Casimirs are quadratic in $k$:~\cite{casimir}
\be
C_2 (k)=\frac{N-1}{N}k(N+k)
\ee

A new area variable, $t$, now replaces $\tau$:
\be
t=\tau\frac{N-1}{N+1} = \lambda{\cal A} \left (1-\frac{1}{N}\right )
\ee
$q_N (y,\tau)$ is replaced by two functions of $z$, $\psi^{(N)}_\pm (z,t)$. 
$\psi^{(N)}_+$ is defined
for $|z|>1$ and $\psi^{(N)}_-$ is defined
for $|z|<1$. The $\psi^{(N)}_\pm$ are analytic in their respective domains. 
\be
\langle \det (z-W)^{-1} \rangle = \psi^{(N)}_\pm (z,t)
\ee
$+$ or $-$ hold, depending on whether $z$ is inside the unit circle or outside it. For $|z|>1$, $\psi^{(N)}_\pm (z,t)$ are given by:
\be
\psi^{(N)}_+ (z,t) =\frac{1}{z^N} \sum_{k=0}^\infty \frac{d_k}{z^k} e^{-\frac{t}{2N} k(N+k)};~~  \psi^{(N)}_- (\frac{1}{z},t)=(-z)^N 
\psi^{(N)}_+ (z,t)
\label{sumeqa}
\ee
At $t=0$ $P_N (W,0)=\delta(W,{\bf 1})$ with respect to the
Haar measure. Therefore,
\be
\psi^{(N)}_\pm (z,0)=\frac{1}{(z-1)^N}
\ee

The linear equation replacing~(\ref{heateq}) is:
\be
\frac{\partial}{\partial t} \psi^{(N)}_\pm (z,t) =
-\frac{1}{2N} \left ( z\frac{\partial}{\partial z} +\frac{N}{2}\right )^2 \psi^{(N)}_\pm (z,t),
\ee
depending on the domain of $z$. $\phi^{(N)} (y,\tau)$ is replaced by: 
\be
\phi^{(N)}_\pm (z,t) =\frac{i}{N} \frac{1}{\psi^{(N)}_\pm (z,t)}\left ( z\frac{\partial}{\partial z} +\frac{N}{2}\right ) \psi^{(N)}_\pm (z,t)
\ee
Explicitly,
\be
\phi^{(N)}_\pm (z,t) =\mp i\left [ \frac{1}{2} +\frac{1}{N} 
\frac{\sum_{k=1}^\infty \, k \, d_k \, z^{\mp k}\, e^{-t\frac{k(N+k)}{2N}}}
{1+\sum_{k=1}^\infty \, d_k \, z^{\mp k} \, e^{-t\frac{k(N+k)}{2N}}}\right ]
\label{sumeqb}
\ee
These functions obey
\be
\frac{1}{2N} \left ( i z\frac{\partial}{\partial z} \right )^2 
\phi^{(N)}_\pm (z,t) + 
\left ( i z\frac{\partial \phi^{(N)}_\pm (z,t)}{\partial z} \right )
\phi^{(N)}_\pm (z,t) =\frac{\partial \phi^{(N)}_\pm (z,t)}{\partial t}
\ee

As before, an exponential substitution leads to Burgers' equation:
\be
z=e^{-iY}
\ee
The map $Y\to z$ takes the real axis into the unit circle, 
the $\Im Y > 0$ half plane into $|z|>1$ and 
the $\Im Y < 0$ half plane into $|z|<1$. Every strip $|\Re Y - 2k \pi | < \pi,~k\in Z$ is mapped onto the entire $z$-plane. In terms of $Y$,
there is only interest in functions periodic under $Y\to Y+2k\pi$, which
define single valued functions of $z$. Viewing the $\phi^{(N)}_\pm$ as functions of $Y$ the nonlinear PDE-s become complex Burgers' equations,
\be
\frac{1}{2N} \frac{\partial^2 \phi^{(N)}_\pm}{\partial Y^2} =
\frac{\partial \phi^{(N)}_\pm}{\partial t}+
\frac{\partial \phi^{(N)}_\pm}{\partial Y}\phi^{(N)}_\pm
\ee
with initial conditions:
\be
\phi^{(N)}_\pm (e^{-iY},0)=\frac{1}{2} \cot\frac{Y}{2}
\ee
The explicit forms of the solutions in terms of sums over $k$~(\ref{sumeqa},\ref{sumeqb}) imply
the following asymptotic behavior at $t\to\infty$:
\be
\phi^{(N)}_\pm (e^{-iY},\infty) =\mp\frac{i}{2}
\ee

\section{Relation to DO.}

At $N=\infty$, $\phi^{(\infty)}_+ (e^{-iY},t)$ is related to the function
$f(A,\alpha)$ of~\cite{duol} by:
\be
f(A,\alpha)=\frac{1}{2\pi}\phi^{(\infty)}_+ (e^{-iY},t)
\ee
with
\be
Y=\alpha,~~~t=\frac{A}{2\pi}
\ee
Therefore, $\phi^{(N)}_+ (e^{-iY},t)$ is one possible extension of the
DO solution to finite $N$. 

At infinite $N$ DO define the eigenvalue density of $W$, $\rho_A (\alpha)$, now for real $\alpha$, by
\be
\rho_A(\alpha) =-2\lim_{\Im \alpha\to 0^+} [\Im f (A,\alpha)]
\ee
Observe that for any $t>0$ the definitions of $\psi^{(N)}_\pm (z,t)$
by sums over $k$ can be analytically extended to all $z$, excepting
$z=0$  for $\psi^{(N)}_+$ and $z=\infty$ for $\psi^{(N)}_-$. In particular, for $t >0$, $\psi^{(N)}_\pm (z,t)$ are well defined for $|z|=1$. The unit circle $|z|=1$ is parametrized by $z=e^{-iy}$ with real
$y$. 

It is easy to check that $\phi_+^{(N)} (e^{-iy},t) +
\phi_-^{(N)} (e^{-iy},t)$ is purely real and 
$\phi_+^{(N)} (e^{-iy},t) -
\phi_-^{(N)} (e^{-iy},t)$ is purely imaginary. This finally leads to
an extension of the DO infinite $N$ eigenvalue density to finite $N$:
\be
\rho^{(N)} (y,t)=\frac{i}{2\pi} [\phi_+^{(N)} (e^{-iy},t) -
\phi_-^{(N)} (e^{-iy},t)]
\ee

The limiting behavior at $t=\infty$ is now seen to be
\be
\rho^{(N)} (y,\infty)=\frac{1}{2\pi},
\ee
and is $N$ independent. The $N$ independent initial condition also 
requires a limiting procedure because singularities appear at $z=1$
when $t$ attains the value 0:
\begin{eqnarray}
&\rho^{(N)} (y,0)=\frac{i}{2\pi} \lim_{t\to 0^+} 
\left \{ \lim_{\epsilon\to 0^+} \left [ \phi_+^{(N)}(e^{-iy+\epsilon},t ) -\phi_-^{(N)} (e^{-iy-\epsilon}, t)\right ] \right \} =\cr
& \frac{i}{2\pi}\lim_{\epsilon\to 0^+} \left [ \frac{1}{2} 
\cot\frac{y-i\epsilon}{2} -\frac{1}{2}\cot\frac{y+i\epsilon}{2} \right ]=\sum_{k=-\infty}^\infty \delta (y-2k \pi )
\end{eqnarray}
The initial condition is also $N$-independent. Thus, the entire $N$-dependence of $\rho^{(N)}$ is contained in the differential equations, more specifically, in their viscous terms.

\section{Equation for $\rho^{(N)}$.}

At infinite $N$ $\rho^{(N)}$ is the Wilson loop matrix eigenvalue density and therefore a more physical object than the average of the
inverse characteristic polynomial. It therefore seems desirable to
derive an equation for $\rho^{(N)}(y,t)$ directly, without the involvement of other functions. 

The equations obeyed by $\phi^{(N)}_\pm (e^{-iy},t)$ have only one  nonlinear term 
\be
\frac{1}{2}\frac{\partial}{\partial y} \left ( \phi^{(N)}_\pm \right )^2
\ee
which hinders superposition. Using 
\be
\left ( \phi^{(N)}_+ \right )^2 -\left ( \phi^{(N)}_- \right )^2=
\left (\phi^{(N)}_+ -\phi^{(N)}_- \right )
\left (\phi^{(N)}_+ +\phi^{(N)}_- \right )
\ee
one could get an equation just for 
$\left (\phi^{(N)}_+ -\phi^{(N)}_- \right )(e^{-iy},t)$ if one expressed 
the sum $\left (\phi^{(N)}_+ +\phi^{(N)}_- \right )(e^{-iy},t)$ in terms of
the difference $\left (\phi^{(N)}_+ -\phi^{(N)}_- \right )(e^{-iy},t)$. This is possible since, for $t>0$, 
$\left (\phi^{(N)}_+ \pm \phi^{(N)}_- \right )(e^{-iy},t)$ 
are the real and imaginary parts of the analytic function 
$\phi^{(N)}_+ (z,t)$ on the curve $|z|=1$. 

The needed device is the Hilbert transform ${\bf H}$, 
mapping functions of a real
variable, $g(y)$ into other functions of a real variable, $({\bf H} g)(y)$:
\be
({\bf H} g )(y)=\lim_{A\to\infty} \frac{P}{\pi} \int_{-A}^A 
\frac{g(x)}{y-x} dx \equiv \frac{P}{\pi} \int_{-\infty}^\infty
\frac{g(x)}{y-x} dx 
\ee
${\bf H}$ is defined with the opposite sign relative to~\cite{baker}. 
For real $y$ and $t>0$ one obtains:
\be
\left ( \phi^{(N)}_+ (e^{-iy},t) \right )^2 
-\left ( \phi^{(N)}_- (e^{-iy},t)\right )^2=
- i\,(2\pi)^2\,\rho^{(N)}(y,t) ({\bf H} \rho^{(N)} ) (y,t)
\ee

The sought after equation follows:
\be
\frac{1}{2N} \frac{\partial ^2 \rho^{(N)} (y,t)}{\partial y^2} =
\frac{\partial \rho^{(N)} (y,t)}{\partial t} + 
\pi\frac{\partial}{\partial y}\left [ \rho^{(N)}(y,t) ({\bf H} \rho^{(N)} ) (y,t)\right ]
\ee

This integro-differential equation is the main result of this letter.
The equation is a particular case of an equation studied in~\cite{matsuno}. The equation has been further investigated in~\cite{baker}. The results of~\cite{baker} were applied in the
context of infinite $N$ to 2D $SU(N)$ YM by Blaizot and Nowak in~\cite{blaizot}. Note that the nonlocal term already contributes
in the inviscid limit. Thus, the result here provides a direct and
minimal extension to finite $N$.

Perhaps the most
interesting property of this integro-differential equation is that turning on the viscosity does not
assure regularity even for smooth initial conditions: one can have
finite time ``blow-ups'', even for finite $N$. In our application we know
that we start from a singular initial condition. Another interesting 
property of this equation is that it admits solutions given by superpositions of pole terms with the entire $t$ dependence given by the
location of the poles in the complex plane. The motion of the poles is
governed by coupled first order differential equations of Calogero type.

More work on the consequences of the above for physics is left for the future.

\section{Integral representations.}

For $t>0$ and $|z|\ne 1$, equations~(\ref{sumeqa}) admit an integral representation which
includes the $t\to 0^+$ initial condition. The basic step is to write
a Gaussian integral representation for the $t$-dependent term:
\be
e^{-\frac{t}{2N} k (N+k) }= e^{\frac{Nt}{8}} \int_{-\infty}^{\infty}
\frac{dx}{\sqrt{2\pi}} e^{-\frac{1}{2} x^2 + i x \sqrt{\frac{t}{N}}(k+\frac{N}{2})}
\ee
So long as $t >0$ and $|z|>1$ 
the sum giving $\psi^{(N)}_+$ can be interchanged with the integral
and after that the sum over $k$ can be performed. Changing variables 
of integration to $u=x\sqrt{\frac{t}{N}}$ produces:
\be
\psi_+^{(N)} (z,t)= e^{\frac{Nt}{8}} \sqrt{\frac{N}{2\pi t}} \int_{-\infty}^{\infty} du \, e^{-\frac{N}{2t} u^2 } \,
\left ( ze^{-i\frac{u}{2}} -e^{i\frac{u}{2}}\right )^{-N}
\label{intone}
\ee
From this, one can immediately get an integral representation for
$\psi_-^{(N)}$. For comparison, the analogue equation in the anti-symmetric case~\cite{first}, for any $z$, is
\be
\langle \det (z-W)\rangle =e^{-\frac{N\tau}{8}} 
\sqrt{\frac{N}{2\pi\tau}} 
\int_{-\infty}^{\infty} du \, e^{-\frac{N}{2\tau} u^2 } \,
\left ( ze^{-\frac{u}{2}} -e^{\frac{u}{2}}\right )^{N}
\label{inttwo}
\ee
Eq.~(\ref{inttwo}) can be obtained from eq.~(\ref{intone}) with
the formal replacements $u\to -iu$ and $\sqrt{N}\to i\sqrt{N}$, 
the latter causing also the replacement 
$t\to\tau$ because we view $\lambda{\cal A}$ as fixed.

These integral representations produce asymptotic expansions in $\frac{1}{N}$ starting from a dominating saddle point 
and make evident the difference in analytic structures in $z$ 
at finite $N$.

\section{Discussion.}

The first choice for an analytic function containing information about
the eigenvalue density of $W$ would be the average resolvent:
\be
\langle R_N (z,t) \rangle =\frac{1}{N} \langle tr (z-W)^{-1} \rangle
\ee
$R_N$ has a natural expansion in $n$-wound loops, $tr W^n$, which
can be converted to an expansion in representations, but at the expense
of linearity; all representations contribute to $\langle R_N \rangle$. 

Here and in~\cite{first}, I focused on the characteristic polynomial
because:
\be
R_N(z,t)=\pm \frac{1}{N} \frac{\partial }{\partial z} \log [\det(z-W)^{\pm 1}]
\label{resdet}
\ee
For simple, properly normalized gauge invariant observables ${\cal O}_i$ 
one has infinite $N$ factorization:
\be
\langle \prod_i {\cal O}_i \rangle = \prod_i 
\langle {\cal O}_i \rangle
\ee
This implies that at infinite $N$ 
\be
\langle f({\cal O} )\rangle = f (\langle {\cal O} \rangle )
\ee
for an arbitrary function $f$ at points where it can be Taylor expanded.
Hence, $\langle R^{(\infty )} (z,t)\rangle $ will be obtained with
either sign choice in~(\ref{resdet}). 

Choosing $+$ in~(\ref{resdet}) and averaging with
respect to $P_N$ produces a sum over $N$ simple poles. However,
the averaged 
eigenvalue density at finite $N$ should be reflected by a cut
running round the unit circle, completely segregating the interior
of the unit circle from its exterior. This indicates that the choice
of $-$ in~(\ref{resdet}) is a more 
appropriate extension of the infinite $N$
eigenvalue density. 

This discussion makes it evident why the extension
to finite $N$ is non-unique. 
The preferred extension would perhaps be  
$\langle R_N(z,t)\rangle$, but the method used for
the characteristic polynomials in this letter fails there, and there is doubt that a simple finite $N$ equation exists. There are other methods
that yield the same result for the characteristic polynomial, and these
methods might extend to $R_N$, starting from
\be
\langle \det \left ( \frac{z_1 -W}{z_2 -W}\right ) \rangle
\ee
and taking $z_1\to z_2$ subsequently. 
I hope to explore this more in the future.

The longer view is to recall that only the 2D problem can reduce to simple
equations, while the main interest is to focus on the large $N$
non-analyticity and its universal smoothing out at finite $N$, 
features which do appear to extend to higher dimensions~\cite{ourjhep, three-d, pcm, lat07, inst}.
There, the main new ingredient is the need to renormalize. 
I think that the most 
convenient object to renormalize is the average characteristic
polynomial, the topic of~\cite{first}, but would not rule
out the average of the inverse characteristic polynomial as deserving
more study in this context either. I hope to be able to provide
a renormalized framework for dealing with the average characteristic
polynomial of a Wilson loop in 3D and 4D some time in the future.

\break

\subsection*{Acknowledgments.}

I acknowledge partial support by the DOE under grant
number DE-FG02-01ER41165 at Rutgers University and by the SAS of Rutgers University.
I note with regret that my research has for a long time been 
deliberately obstructed by my high energy colleagues at Rutgers.  
An ongoing collaboration on related topics
with R. Narayanan is gratefully acknowledged.

\end{document}